\documentclass[titlepage,twoside,12pt]{article}
\usepackage{amssymb}
\usepackage{amsfonts}
\begin{document}

{\bf \Large Theories of Physics and Impossibilities} \\ \\

{\bf Elem\'{e}r E Rosinger} \\
Department of Mathematics \\
and Applied Mathematics \\
University of Pretoria \\
Pretoria \\
0002 South Africa \\
eerosinger@hotmail.com \\ \\

{\bf Abstract} \\

The role of {\it impossibilities} in theories of Physics is mentioned and a recent result is
recalled in which Quantum Mechanics is characterized by three information-theoretic
impossibilities. The inconvenience of the {\it asymmetries} established by such
impossibilities is pointed out. \\ \\

{\bf 1. Preliminary Remarks} \\

Theories of Physics were seen by Einstein as falling into two main categories, [1]. \\
Some are {\it constructive} as :

\begin{quote}

"they attempt to build up a picture of the more complex phenomena out of the materials of a
relatively simple formal scheme from which they start out. Thus the Kinetic Theory of Gases
seeks to reduce mechanical, thermal and diffusional processes to movements of molecules ...".

\end{quote}

Other theories can be seen as {\it principle theories}, since :

\begin{quote}

"these employ the analytic, not synthetic method. The elements which form their basis and
starting point are not hypothetically constructed but empirically discovered ones, general
characteristics of natural processes, principles that give rise to mathematically formulated
criteria which the separate processes or the theoretical representations of them have to
satisfy. Thus the science of Thermodynamics seeks the analytical means to deduce necessary
conditions, which separate events have to satisfy, from the universally experienced fact that
perpetuum motion is impossible."

\end{quote}

And in Einstein's view the Theory of Relativity, for instance, belongs to the second above
category. \\

These second category theories of Physics recall Euclidean Geometry which had impressed
Einstein himself during his school years. Indeed, one starts from certain empirically evident
principles, and then based on them, constructs the whole theory by using logical deductions. \\

Another most intriguing remark of Einstein relates to the foundational role of {\it
impossibilities} in certain theories of Physics, [2] :

\begin{quote}

"The totality of physical phenomena is of such a character that it gives no basis for the
introduction of the concept of 'absolute motion', or shorter but less precise : There is no
absolute motion. It might seem that our insight would gain little from such a negative
statement. In reality, however, it is a strong restriction for the conceivable laws of nature.
In this sense there exists an analogy between the Theory of Relativity and Thermodynamics. The
latter, too, is based on a negative statement : 'There exists no perpetuum mobile.' "

\end{quote}

In fact, Special Relativity is based solely on two impossibilities :

\begin{itemize}

\item there is no absolute motion,

\item no physical entity can move faster than light in vacuum.

\end{itemize}

The remarkable fact is that in Physics, within the second category theories, one can start
with principles expressed by very simple {\it impossibilities}, like for instance those
mentioned above. \\ \\

{\bf 2. Three Impossibilities as the Foundation of \\
\hspace*{0.4cm} Quantum Mechanics} \\

Recently in [3], see also [4, 5], it was shown that Quantum Mechanics can be {\it
characterized} by the following three information-theoretic {\it impossibilities} :

\begin{itemize}

\item the impossibility of superluminal information transfer between two physical systems by
performing measurements on one of them,

\item the impossibility of perfectly broadcasting the information contained in an unknown
physical state, which impossibility for pure states amounts to "no cloning",

\item the impossibility of communicating information so as to implement a "bit commitment"
protocal with unconditional security.

\end{itemize}

\bigskip

{\bf 3. Comments} \\

One may, of course, think that in the case of second category theories of Physics, namely,
those called "principle theories" by Einstein, it is rather trivial to formulate the
respective principles as impossibilities. Indeed, any principle, say, "P", and not only in
Physics, can be stated equivalently by its double negation "non-non-P". And then this
equivalent form is nothing else but stating as principle the impossibility of "non-P". \\

From the above examples, however, it is clear that the respective impossibilities are {\it not}
of that trivial form. Indeed, as they are given, none of them is a double negation, but only a
simple, one time negation. And as such, they describe definite binary choices concerning
fundamental properties of the whole of their corresponding realms of physical situations,
choices in which, a priori and on purely logical grounds, both alternatives may appear to be
possible. Furthermore, and quite importantly, they describe binary choices in which one of the
alternatives may appear to be particularly {\it convenient}, yet it is precisely that
alternative which ends up being denied in principle. \\

In other words, these impossibility principles are establishing {\it asymmetries} in their
respective realms, and do so in ways which appear to be inconvenient. \\

For instance, a priori, it may not be clear whether there is, or on the contrary, there is no
perpetuum mobile. And needless to say, it would be so much more convenient for us if there
were any at all. Yet the principle adopted denies the existence of even one single perpetuum
mobile. \\
A similar situation happens with the limitation given by the velocity of light in vacuum, or
for that matter, with the above three impossibilities which characterize Quantum Mechanics. \\

\end{document}